\documentstyle[aps,preprint]{revtex}

\tightenlines 
\begin{document}

\title{Critical behavior of a one-dimensional fixed-energy stochastic  
sandpile} 
     
\author{Ronald Dickman$^1$, Mikko Alava$^2$, Miguel A. Mu\~noz$^{3}$, 
Jarkko Peltola$^2$, Alessandro Vespignani$^{4}$, and Stefano Zapperi$^{5}$} 
\address{ 
$^1$ Departamento de F\'{\i}sica, ICEx, 
Universidade Federal de Minas Gerais, 
Caixa Postal 702, 
30161-970 Belo Horizonte, MG, Brazil\\ 
$^2$ Helsinki University of Technology, Laboratory of Physics,  
HUT-02105 Finland \\ 
$^3$ Institute {\it Carlos I} for Theoretical and Computational Physics\\ 
and Departamento de Electromagnetismo y F{\'\i}sica de la Materia\\ 
18071 Granada, Spain.\\ 
$^4$ The Abdus Salam International Centre for Theoretical Physics (ICTP) 
P.O. Box 586, 34100 Trieste, Italy\\ 
$^5$ INFM, Dipartimento di Fisica, E. Fermi, 
Universit\'a de Roma ``La Sapienza", P.le A. Moro 2, 
00185 Roma, Italy \\ 
} 
\date{\today} 
 
\maketitle 
\begin{abstract} 
We study a one-dimensional fixed-energy version
(that is, with no input or loss of particles),
of Manna's stochastic sandpile model. 
The system has a  continuous transition to an absorbing 
state at a critical value $\zeta_c$ of the particle density.   
Critical exponents are obtained from 
extensive simulations, which treat both stationary 
and transient properties.   In contrast with other one-dimensional 
sandpiles, the model appears to exhibit finite-size scaling,  though 
anomalies exist in the scaling of relaxation 
times and in the approach to the stationary state. 
The latter appear to depend strongly on the nature of the
initial configuration.
The critical exponents differ from those expected at a  
linear interface depinning transition in a medium with point 
disorder, and from those of directed percolation.   

\end{abstract} 
 
 
 
\date{\today} 
 
\section{Introduction} 
 
Sandpile models are the prime example of self-organized
criticality (SOC), or scale-invariance in the apparent 
absence of tuning
parameters \cite{btw,ggrin,bjp,AIP}.   
SOC in a slowly driven sandpile has an analog in the 
absorbing-state phase transition of the corresponding nondriven  
or fixed-energy sandpile (FES) \cite{bjp,AIP,tb88,vz,dvz,vdmz}.   
While most studies of sandpiles have probed 
the slow-driving limit (addition and loss of sand 
grains at an infinitesimal rate), there 
is great interest in understanding the scaling 
properties of FES models as  
well \cite{dvz,mdvz,chessa,carlson}.  
One may hope that these systems, which are isolated 
and translation-invariant, will prove easier to characterize than 
driven sandpiles, which have eluded a rather massive theoretical 
and computational effort  
to obtain precise critical exponents or assign universality classes.   
In this paper we present extensive numerical results for a particularly 
simple one-dimensional FES. Analogous studies of higher
dimensional systems can be found in \cite{dvz,vdmz,fes2} (A review of 
these results and related models can be found in Refs.~\cite{bjp,AIP}). 
 
One of the central features of sandpile models 
is the presence of a conserved field, 
which may be thought of as the density of sand grains, or, in our 
nomenclature, the energy density.   This field is conserved locally as 
well as globally,  
and couples to the activity density, which is the order parameter.  The 
configuration evolves through a series of ``toppling" events, i.e.,  
redistributions of energy from an active 
site to its neighbors, which may be either deterministic or stochastic.   
The well known 
Bak-Tang-Wiesenfeld (BTW) sandpile has a deterministic toppling rule, 
and is also Abelian (the final absorbing configuration is independent  
of the order of 
toppling), which allows many stationary-state properties 
of the driven sandpile to be found 
exactly \cite{dhar99,priez}.  
A less desirable aspect of the deterministic  
dynamics, on the other hand, is that in the steady-state only  
a small subset of the possible configurations (determined by the 
initial state) are visited \cite{dhar99}. 
This leads to strong nonergodic effects in the FES version of the
BTW automaton \cite{fes2}. 
Here we study a one-dimensional stochastic FES sandpile to elucidate the 
consequences of stochastic rules.  
 
In our model, a variant of the Manna 
sandpile \cite{manna,manna2},  the pair of particles 
liberated when a site topples move {\it independently,} 
and with equal probability, to the nearest neighbors on the left 
or right.  We may therefore think of the particles 
as independent random walkers.  (There is no 
restriction on the number of walkers at a given site.)   
While the walkers are independent 
as far as their hopping direction is concerned, their mobility does involve  
an interaction: isolated particles cannot move, but if 
two or more particles occupy the same site, the site may topple.  
Thus for densities $\zeta = N/L \leq 1$ ($N$ is the  
number of walkers on a ring  of $L$ sites),
absorbing configurations exist, in which all walkers are paralyzed.  
But since the fraction of absorbing configurations 
vanishes as $\zeta \to 1$, it is reasonable to expect a 
phase transition from an absorbing to an active phase at 
some $\zeta_c <1$.  Simulations bear this out, and 
show that there is a continuous transition
at $\zeta_c \simeq 0.9488$.
 
Our goal in the present paper is to determine the critical 
behavior of the model. To this end, we analyze the static behavior 
and the approach to the steady state in the language of absorbing-state 
phase transitions, studying the order parameter,  
its fluctuations, and temporal correlations. 
We also investigate the cumulative local activity, whose 
dynamics obeys a linear diffusion equation in the presence of 
an effective quenched noise term that reflects the stochastic   
redistribution rules, 
the Linear Interface Model (LIM) \cite{midd,pacz,lau,ala}.  
Our main finding is that the one-dimensional Manna model defines a
universality class different both from that of directed percolation,
and of the linear interface model. The existence of 
long-range correlations that arise from
a conservation law is believed to be at the basis of this
new universality class \cite{dvz,vdmz,fes2,alepp}.

The paper is structured as follows.  
In Sec. II we define the model and our simulation algorithm.   
Numerical results are analyzed, in the contexts of absorbing-state phase transitions
and of driven interfaces, in Sec. III.  In Sec. IV we summarize and discuss our findings. 
 
\section{Model} 
 
The model is defined on a one-dimensional 
lattice with periodic boundaries. The 
configuration is specified by the energy (or number of walkers) 
$z_i = 0, 1, 2,...$ at each site; sites with $z_i \geq 2$ are said to be 
{\it active}.  A Markovian dynamics is defined by the toppling rate, which is 
unity for all active sites, and zero for sites with $z_i < 2$.  
When a site $i$ topples, $z_i \to z_i - 2$ and the 
two particles liberated move {\it independently} to randomly  
chosen nearest neighbors $j$ and $j'$ ($j$, $j' \in \{i+1,i-1\}$).   
(Thus $j=j'$ with probability $1/2$.) 
In most cases we use random sequential dynamics: the next site to topple is  
chosen at random from a list of active sites, which must be  
updated following each toppling event.  The time increment associated with  
each toppling is $\Delta t = 1/N_A$, where $N_A$ is the number of active  
sites just prior to the toppling. 
In this way, $\langle N_A \rangle$ 
sites on average topple per unit time, just as they would in a 
simultaneously updated version of the model, in which all active 
sites at time $t_n$ are toppled simultaneously and $\Delta t \equiv 1$. 
We expect the two dynamics to be equivalent 
insofar as scaling properties are concerned, and note that the 
latter choice was used in some of the interface representation studies
discussed below. 
 
In most of our simulations, the initial condition is generated by 
distributing $\zeta L$ particles randomly among the $L$ sites, 
yielding an initial (product) distribution that is spatially homogeneous 
and uncorrelated. 
Once the particles have been placed, the dynamics begins  
(we verified that allowing some 
toppling events {\it during} the insertion 
phase has no effect on the stationary properties). 
The condition of having at least one active site 
in the initial configuration is trivially satisfied on large lattices, 
for the $\zeta$ values of interest, i.e., close to the critical value. 
In fact, for large $L$, the initial height at a given site is essentially a  
Poisson random variable, and the probability of having no active sites 
decreases exponentially with the lattice size.  
It is worth remarking that while the initial 
conditions are statistically 
homogeneous, the energy density is not perfectly smooth. 
For $ 1 \ll l \ll L$, the 
energy density on a set of $l^d$ sites is essentially a 
Gaussian random variable with mean $\zeta$ and variance $\sim l^{-d}$. 
Such initial fluctuations relax on a time scale $\sim l^{z_D}$ with
$z_D \geq 2$.  That is, the exponent governing diffusive relaxation must
be at least as large as for independent random walkers.  (Recall that
in our model regions of lower density will exhibit less activity, hence
a slower redistribution of particles.)
The initial value of the critical-site density, $\rho_c$ 
(the density of sites with $z_i = 1$), moreover, is generally 
far from its stationary value, complicating  
relaxation to the steady state. 
These observations are pertinent to the studies of modified initial 
conditions reported in Sec. IIId.

If after some time the system falls into a configuration with  
no active sites, the dynamics is permanently frozen, i.e., the  
system has reached an absorbing configuration. 
We shall see that as we vary $\zeta$,  
the model exhibits a phase  
transition separating an absorbing phase (in which   
the system always falls into an absorbing configuration), from an active 
phase possessing sustained activity.

\section {Simulation Results} 
 
\subsection{Absorbing-state phase transition} 

The presence of an absorbing-state phase transition is a 
general feature of FES models\cite{dvz,vdmz,fes2}.  
In analogy with standard analyses of such transitions, it is possible to
derive a set of Langevin equations describing their critical behavior
in high dimensions. 
Following Ref.\cite{fes2},
we can write the dynamical equations for the density of active sites 
$\rho_a$ and the energy density $\zeta$ as  
\begin{equation}
\partial_t \rho_a=D\nabla^2\rho_a +\mu\rho_a -b\rho_a^2 + w\zeta\rho_a
+\rho_a^{1/2}\eta
\label{eq:1}
\end{equation}
\begin{equation}
\partial_t\zeta= \lambda\nabla^2\rho_a ,
\label{eq:2}
\end{equation}
where $\mu,b,w$ and $\lambda$ are coupling constants and $\eta$ is a 
Gaussian noise. Eqs.~(\ref{eq:1}-\ref{eq:2}) closely resemble 
the Langevin field theory of directed percolation (DP), apart from 
an additional coupling between the activity
and the energy field, which is conserved by the dynamics. 
Due to this coupling, the FES critical  behavior is not 
in the DP universality class, but belongs to a new universality class
embracing all systems with multiple absorbing states in which
the order parameter is coupled to a static conserved field \cite{alepp,pv}.
(By `static' we mean that this field is frozen in regions where the
order parameter is null.)
The field theory for a reaction-diffusion model in this universality 
class has been derived exactly; it reproduces
in some limits Eqs.~(\ref{eq:1}-\ref{eq:2}) \cite{pv}. 
It is thus interesting to study the critical behavior 
in $d=1$, to see if it is possible
to generalize the known results for fixed-energy sandpiles \cite{fes2}  
to one-dimensional systems.  

We simulated the model using system sizes ranging from 
$L = 100$ to about $10^4$ sites. 
(Since $\zeta = N/L$ with $N$ an integer, we are obliged to use different  
sets of $L$ values to study different energy densities $\zeta$.) 
In stationary-state simulations, we collect data over 
an interval of $t_m$ time units, following a relaxation period of $t_r$.  
For small system sizes, $t_m$ and $t_r$ are of the order of 
10$^3$, but for our largest systems 
($L \simeq 10^4$) we used $t_r \geq 5 \times 10^6$ and 
$t_m = 2.5 \times 10^6$. 
We verified that our results show no systematic variation with time  
for $t > t_r$. A run consists of $N_s$ independent trials, 
each with a different initial configuration, with   
$N_s$ ranging from $2 \times 10^5$ 
for $L=100$, to 500 or 1000 for $L \simeq 10^4$. 
In practice $t_m$ is limited because 
for $\zeta \simeq \zeta_c$, the survival probability 
decays sensibly over this time scale; in some cases 
only about 25\% of the trials survive to time $t_r + t_m$.

Fig.~1 shows the overall dependence of the stationary active-site density  
as a function of $\zeta$; the points represent extrapolations of results  
for $L = 100 $ - $5000$ to the $L \to \infty $ limit.
The data indicate a  
continuous transition from an absorbing state ($\rho_a$ = 0) to an active  
one at $\zeta_c$ in the vicinity of 0.95. 
 
Our first task is to locate the critical value $\zeta_c$. 
We studied the stationary active-site density, $\rho_a$, 
and its second moment $\rho_a^2$, 
anticipating that as in other absorbing-state phase transitions, 
the active-site density (i.e., the order parameter) 
will obey finite-size scaling \cite{fss}: 
 
\begin{equation} 
\overline{\rho_a} (\Delta,L) =  
L^{-\beta/\nu_{\perp}} {\cal R}  
(L^{1/\nu_{\perp}} \Delta) \;, 
\label{actfss} 
\end{equation} 
 
\noindent where $\Delta \equiv \zeta - \zeta_c$, and 
${\cal R}$ is a scaling function with ${\cal R}(x) \sim x^{\beta}$ 
for large $x$,  
since for $L \gg \xi\sim \Delta^{-\nu_{\perp}}$ we expect 
$\overline{\rho_a}\sim \Delta^\beta$ (here $\xi$ is the correlation length).   
When $\Delta=0$ we have that $\overline{\rho_a} (0,L)  
\sim L^{-\beta/\nu_{\perp}}$. For $\Delta>0$, by contrast, 
$\overline{\rho_a}$ approaches a stationary value, while for  
$\Delta<0$ it falls off as $L^{-d}$. 
Thus in a double-logarithmic plot of $\overline{\rho_a}$ versus $L$  
(see Fig.~2), 
supercritical values ($\Delta > 0$) are characterized by 
an upward curvature, while for $\Delta < 0$ the graph curves downward. 
Using this criterion (specifically, zero curvature in the data for  
$L \geq 1000$), 
we find $\zeta_c = 0.94887(7)$, with the uncertainty reflecting the 
scatter in our numerical results for the curvature (see Fig.~2, inset). 
The associated exponent ratio is $\beta/\nu_{\perp} = 0.235(11)$. 
A similar analysis of the data for $\rho_a^2$ yields 
$\zeta_c = 0.94883(5)$ with an exponent of $2 \beta/\nu_{\perp} = 0.483(18)$. 
We therefore adopt the estimates 
$\zeta_c = 0.94885(7)$ and $\beta/\nu_{\perp} = 0.239(11)$. 
 
We measured the 
autocorrelation function for the number of active sites $N_a$, 
 
\begin{equation} 
C(t) = \frac{\langle N_a (t_0+t) N_a (t_0) \rangle - \langle N_a \rangle^2} 
       {\langle N_a^2 \rangle - \langle N_a\rangle ^2} 
\label{ct} 
\end{equation} 
in the stationary state.  To obtain clean results for $C(t)$ we study  
surviving trials in relatively long runs (from $t_m = 2 \times 10^5$ for 
$L=625$, to $5 \times 10^6$ for $L=10^4$; this obliges us to reduce our 
sample to 200 surviving trials for $L \leq 2500$ and 100 surviving trials 
for $L \geq 5000$).  Results for $\zeta = 0.9488$ are shown in Fig. 3: 
C(t) decreases monotonically, but does not follow 
a simple exponential decay.  To study the dependence of the relaxation 
time on system size, we determine the temporal rescaling factor $r$ required 
to obtain a data collapse between $C(t;L/2)$ and $C(t/r;L)$.  
A good collapse is 
possible (see Fig. 3), but the rescaling factor depends on $L$; 
for $L= 2^n \cdot 625$, we use $t^* = t/r^n$ with 
$r=2.93$, 2.91, and 2.89 for $n = 1$, 2, and 3, respectively. 
The rescaling factor $r$ appears to approach a limiting ($L \to \infty$)  
value of 2.80(5), corresponding to a relaxation time that scales as  
$\tau \sim L^{\nu_{||}/\nu_{\perp}}$ 
with $\nu_{||}/\nu_{\perp} \equiv z = \ln r / \ln 2 = 1.49(3)$. 
We also studied the half-life $\tau_H$ (the time for the survival probability 
to decrease by a factor of 1/2 in the stationary state) at $\zeta_c$.   
This relaxation time exhibits a clean power-law dependence on system size, 
$\tau_H \sim L^z$, with $z = 1.42(1)$. 
 
Next we examine the stationary scaling of the order parameter 
away from the critical point. 
We determined the stationary active-site density 
$\overline {\rho_a} (\zeta,L)$, for $\zeta $ in the vicinity 
of $\zeta_c= 0.94885$, and for 
system sizes $L = 100$ - 5000. 
We analyze these data using the finite-size scaling form  
of Eq. (\ref{actfss}), which implies that a plot of 
$L^{\beta/\nu_{\perp}} \overline{\rho_a} (\Delta,L) $ versus 
$L^{1/\nu_{\perp}} \Delta $ should exhibit a data collapse.  We shift 
each data set (in a log-log plot of 
$L^{\beta/\nu_{\perp}} \rho_a $ versus  $L^{1/\nu_{\perp}} \Delta$), 
vertically by $(\beta/\nu_{\perp}) \ln L $, using 
$\beta/\nu_{\perp} = 0.239$ as found above, and determine the  
horizontal shift $S(L)$ required for data collapse. 
These are found to follow the relation $S(L) = \nu_{\perp}^{-1} \ln L$,  
with $\nu_{\perp}^{-1} = 0.553(3)$.   
That these values yield an excellent data collapse is evident from Fig. 4. 
The slope of the scaling plot (linear regression using the 25 points 
with $\ln (L^{1/\nu_{\perp}} \Delta) > -0.5$), 
yields $\beta = 0.410(4)$.  This is somewhat smaller than, 
but consistent with, 
the estimate $\beta = 0.43(2)$ obtained by combining 
$\nu_{\perp}^{-1} = 0.553(3)$ 
and $\beta/\nu_{\perp} = 0.239(11)$.  We adopt 
$\beta = 0.42(2)$ as our final estimate.  
 
\subsection{Interface Representation} 
 
Interface representations are useful for studying both  
sandpiles \cite{midd,pacz,lau,ala} and particle systems such as  
the contact process \cite{rdmam}.  Using such a representation, 
the activity history of the sandpile can be 
described in terms of an equation of motion that is 
an exact analog of a driven interface advancing in a
quenched disordered medium. Our one-dimensional 
case is an example of a larger class of models, 
for which a mapping can be provided from the sandpile 
to an interface field \cite{midd,pacz,lau,ala}.
Such a description can be useful, for example, 
in understanding the role of microscopic rules and in  
defining novel quantities for study. 
 
The mapping is constructed as follows.  Let the interface
height variable $H_i (t)$ count the number of  
topplings at site $i$ up to time $t$. The dynamics of $H_i (t)$ 
follows from the observation that the toppling of a site depends (through its
height $z_i$) on the 
number of particles it has received, and 
on the number of times it has toppled.
Each time a neighbor of site $i$ topples, site $i$ gains one, two, or 
no particles, with probabilities of 1/2, 1/4 and 1/4, respectively. 
This allows a `projection trick', in which the dynamics 
of $H_i(t)$ is constructed from particle counting \cite{ala,fes2}. 
Topplings at the sites neighboring $i$ give rise to an  
average incoming flux $\bar{n}_i^{in}$ at site $i$, 
plus a fluctuating part $\delta n_i^{in}$, which records, in an exact 
manner, the random choices made for each particle.  
The average flux minus the number of topplings at $i$ gives rise to 
a discrete Laplacian, while the fluctuating part can be 
mapped to a noise term $\tau(i,H)$, which can be evaluated for 
each toppling. Following these prescriptions, 
we find a discrete interface 
equation with quenched noise, which reproduces 
the sandpile activity \cite{ala}: 
 
\begin{equation} 
\frac {\partial H_i}{\partial t} = 
\Theta[ \nabla^2 H_i + F_i - 1 + \tau(i,H)] 
\equiv \Theta(f_i) 
\end{equation} 
where $\Theta$ is the step function, so that $H_i$ is nondecreasing. 
Since $H_i(t)$ is restricted to integer values the time derivative 
should be interpreted as the rate of transitions from $H$ to $H+1$. 
The `force' $f_i$ at site $i$ consists of the following contributions. 
The (discrete) Laplacian term arises, as noted 
above, from the balance between the average number of particles 
site $i$ gains from, and loses to, 
its neighbors in toppling events.  
The `columnar' noise term $F_i$ represents the number of particles
initiallly at site $i$,  
while $\tau(i,H)$ is the fluctuation part of the particle flux associated with   
the $H$-th toppling of site $i$.  
The step-function character of the interface equation  
is discussed in greater detail in Ref.~\cite{ala}. 
 
The noise-term $\tau(i,H)$ is a {\it quenched variable} that arises from 
a mapping of the `annealed' disorder of the original sandpile. Note 
that one could also consider the random choices of the particles' 
destinations at each toppling as a quenched, random landscape that 
is chosen in advance for each sandpile configuration. It is then
easier to see that this maps into a quenched disorder term for 
the associated interface equation. 
This noise term is particle-conserving, 
and it has at each $i$ a random-walk-like correlator 
$\left< [\tau(H+\Delta H,i) - \tau(H,i)]^2 \right> \sim \Delta H$. 
If these correlations could be neglected, the scaling of the interface  
would follow that of the linear interface model in a quenched random 
field, as appears to be the case in two- and three-dimensional systems
\cite{fes2}. One should note, however,
that long range-correlations in the noise can change the universality
class, as has been shown numerically for another interface model \cite{sv99}.
The possible LIM universality classes are still under debate \cite{cgl00}.
 
In the interface description the model undergoes a depinning transition 
at the critical force $F_c \equiv \zeta_c$ \cite{barabasi,hhz}. 
At this point power-law correlations develop in the history of topplings. 
The interface behavior, assuming simple scaling, is described 
by two exponents: the roughness exponent $\alpha$, and the  
early-time exponent $ \beta_W$.  Introducing the width $W$ as usual,   
\begin{equation}
 W^2(t,L) = \left< [H_i(t) - \bar{H}(t)]^2 \right>,
\end{equation} 
(here $\bar{H}(t)$ is the mean height), these exponents are defined via 
\begin{equation} 
 W^2 (t,L)  \sim 
 \left\{ 
\begin{array}{ll}
 t^{2 \beta_W} ~~ &  t \ll t_{\times}   \\ 
 L^{2 \alpha}  ~~ &  t \gg t_{\times}  
 \end{array} 
\right., 
\label{w2scal} 
\end{equation} 
where the crossover time $t_{\times} \sim L^z$. 
Assuming that simple scaling holds 
and that the correlations in the interface can be described by  
a single length scale, we have the exponent relation  
$\beta_W z = \alpha$. 
 
This scaling picture, familiar from the study of surface growth,
was recently shown to apply in the case of a simple absorning-state
phase transition \cite{rdmam}.
For the one-dimensional Manna model the situation is 
complicated by several factors. First, the noise appearing in 
the interface description has two components, $\tau(i,H)$ 
and $F_i$. The interface behavior will therefore exhibit a 
crossover from a regime dominated by the initial configuration 
(reflected in $F_i$) to a randomness-dominated 
regime. This effect also appears in higher dimensions, but 
in $d=1$, due to the meager phase space, relaxation is much slower
and transient effects may be much more severe.
Note that the $H$-independent terms on the r.h.s. of the interface equation can be understood
as an initial height profile, which then relaxes to the
asymptotic rough state \cite{ala}. It is easy to see that by varying
such an initial condition different transients can be generated.

A problem with one-dimensional models is that the two-point
correlation function of the surface roughness scales with
a different exponent, $\alpha_{loc}$, which for fundamental
reasons is limited to $\alpha_{loc} \leq 1$.  The exponent
$\alpha$ can attain a larger value, for example $\alpha= 1.25$
for the one-dimensional linear interface model \cite{Lesch}.
These exponents are related via $\alpha = \alpha_{loc} + \kappa$,
where $\kappa$ measures the divergence of the height difference
between neighboring sites with $L$. 
In the corresponding 1-d linear interface model, this is termed
`anomalous scaling' \cite{Lesch,Lopez_97}.
This means that as $t \to \infty$, the typical
height difference between neighboring sites increases without limit.
Since larger systems have a greater lifetime, this
has implications for the roughness as measured by
$W^2_{sat} \equiv \lim_{t \rightarrow t_c} W^2(t,L)$, with
$t_c$ the time at which
the absorbing state with no activity is reached, i.e., the lifetime.
The saturation width $W_{sat}$ scales as $L^\alpha$,
with $\alpha$ related to $\beta_W$ and $z$ as above.
In models exhibiting an absorbing state, such as the contact
process or a FES, the width saturates only because all activity
eventually ceases; the width in {\it surviving} trials does not saturate.
This is in marked contrast to the behavior of interface models,
in which the width saturates due to the Laplacian term representing
surface tension. (Continuum descriptions of absorbing-state phase transitions,
and their associated interface representations, likewisw include a Laplacian
term, but the noise driving the interface grows without limit, so saturation is not
required while there is activity \cite{rdmam}.)

Finally, in absorbing-state models, interface scaling appears to be 
strongly linked to the approach to the stationary state.  In a model with
simple scaling (i.e., unique diverging length and time scales, and no
conserved quantities), the growth exponent $\beta_W$ is related to the
critical exponent $\theta $ governing the initial decay of activity 
via $\beta_W + \theta = 1$ \cite{rdmam}.  In the
present case relaxation is complicated by effects that may mimic (for
a certain time) columnar disorder, and slowly relaxing perturbations
(the long-wavelength density fluctuations mentioned in Sec. II).  

In simulations
we first studied the time-dependent mean interface width $W^2(t,L)$ 
in systems of size $L= 1253$, 2506, 5012, 10024, and 20048, 
at the critical point $\zeta = 0.94892$.   The dependence of the 
saturation width on system system size yields  $\alpha = 1.42(1)$.
We may then attempt to collapse the data for $W^2 (t,L)$
using this exponent, and varying $z$ to obtain the best collapse; 
in this way we find $z = 1.65(2)$.
The resulting
scaling plot (Fig. 5), of $\tilde{W}^2 \equiv W^2/L^{2 \alpha}$ versus 
$\tilde{t} \equiv t/L^z$, shows a good collapse, and an apparent power-law
growth in the roughness, following an initial transient.
From the scaling relation 
$\beta_W = \alpha/z$ we obtain $\beta_W = 0.863(13)$, while a
direct fit to the time-dependent width data yields $\beta_W = 0.87(2)$.

For comparison, an independent series of studies at $\zeta = 0.9490$ 
were performed to determine the lifetime $t_c$ and saturation 
width $W_c$ for $L = 400$ to $L = 6400$. 
Power-law fits to these data yield essentially consistent results, i.e., 
$\alpha = 1.48(2)$, 
$\beta_W = 0.86(2)$ and $z =1.70(3)$.  Fig. 6 shows a clear power-law 
dependence of the saturation width on the lifetime in {\it individual runs}.  
This indicates that a normal Family-Vicsek-style scaling plot of 
$W(t,L)$ as in Fig. 5 is problematic, since as noted above the width in any surviving
run does not saturate as would be the case in normal interface 
models, and the 
saturation width $W \sim w_c$ will therefore depend on how long 
the particular configuration stays alive. 
 
Similarly to the case of the linear interface model, 
we find that there is an independent {\it local} roughness  
exponent $\alpha_{loc}$ that describes the two-point $k$-th 
order height-height correlation function 
$G_k (r) = \left<|H_{i+r} - H_i|^k \right> \sim r^{k\alpha_{loc}}$ 
for $r < \xi(t) \sim t^{1/z}$. 
We find $\alpha_{loc} = 0.59(3)$ for 
$k=1\dots8$. This shows that the  
interface is simply self-affine, not multifractal.  
We note, however, that the interface exponents 
of the Manna sandpile are {\it not} those of the 
one-dimensional LIM. This is most likely due to the fact that, perhaps 
differently from the two- and higher-dimensional cases,
here it is important that the noise term  
{\it increases} in strength with the propagation of the 
interface, or with sustained activity. This arises from 
the fact that the fluctuations in the number 
of grains received at each site, reflected in $\tau(H,i)$, 
follow random-walk characteristics.  
Thus the anomaly 
exponent $\kappa$ indicates an even stronger dependence of the 
step height on $L$ at saturation than in the LIM. 
The same is also true if the step height is considered 
as a function of time for $t < t_c$. 
We find that 
$\kappa_{Manna} \sim 0.82 \sim \kappa_{LIM} + 0.5$. 
Due to this time- or height-dependence of $\tau$ there 
is a direct coupling of $t_c$ and $w_c$ with the final 
average height, $H_c$ (see Fig. 7). As discussed above, this
result follows, in the interface picture, from the fact that
the noise strength increases with time. In the LIM,
$t_c$ and $w_c$ are not correlated on a sample-to-sample basis.

\subsection{Initial Relaxation} 
 
A comparison of our results for stationary correlations ($z \simeq 1.45$) 
with those for the interface ($z \simeq 1.7$) suggests 
that the relaxation to the steady state and 
relaxation of fluctuations {\it within} the stationary regime show 
distinct scaling properties.  
This is supported by our results for the relaxation of the 
active-site density.  
At the critical point of a simple absorbing-state model such as 
the contact process, the activity density $\rho_a$ exhibits an initial 
power-law decay, $\rho_a \sim t^{-\theta}$, followed by a crossover to  
the quasi-stationary value $\rho_a \sim L^{-\beta/\nu_{\perp}}$ \cite{marro}. 
As noted above the growth exponent is related to the activity-decay 
exponent via $\theta + \beta_W = 1$ 
if only one timescale is present \cite{rdmam}.   
A plot of $L^{\beta/\nu_{\perp}} \rho_a (t)$ 
versus $t/L^z$ yields a data collapse to a 
scaling function that is independent of $L$. 
In the present case (Fig. 8),
we see that the collapse is imperfect, and that the form 
of $\rho_a (t)$ changes with $L$.  Here $z$ was chosen so as to optimize 
the collapse at long times, yielding $z=1.75(3)$.
For large systems, 
the active-site density exhibits three distinct regimes 
before reaching the quasi-stationary state: 
an initial power-law decay (I), followed by a crossover 
to a slower power-law regime (II),  
and finally a rapid decay 
(III) to the stationary state.   
For $L=20048$, the exponents associated with regimes I and II are 
0.163 and 0.144, respectively. 
While the latter exponent is in reasonable 
agreement with the scaling relation 
$\beta_W + \theta = 1$, it is clear that relaxation   
to the stationary state is more complicated for the sandpile 
than for, say, the contact process, which presents a unique power-law regime. 
A qualitative explanation may be found in the interface representation:
the initial dynamics is dominated by relaxation of the initial
grain profile $z_i$, which in the interface language means that at
short times the columnar noise, $F_i$, dominates.

Thus another facet of the relaxation process 
is the approach of the mean height 
to its global value, $\zeta$, at a site with initial height $z(0)$. 
In Fig. 9 we plot the mean height $\langle z(t)|z(0)\rangle$  
in at system of 1400 sites at $\zeta_c$, averaged 
over 2000 trials.  
The inset shows that the asymptotic approach to $\zeta$ is    
approximately power-law,   
$|\langle z(t)|z(0)\rangle - \zeta | \sim t^{-\phi} $, 
with $\phi = \simeq 0.46$, 0.45, 0.47, 0.50, and 0.53 for  
$z(0) = 0$, 1, 2, 3 and 4, respectively. 
All of these exponents are close to 
$\phi = 1/2$, the value expected for 
uncorrelated diffusion. 
 
\subsection{Effects of initial preparation}

In light of the complicated pattern of relaxation noted in the preceding 
subsection, it is well to recall that initial-decay studies of simple 
models such as the contact process employ a uniform, featureless initial 
configuration, far from the absorbing state.  This is usually arranged by 
setting the the activity density to unity, an option that is not available 
in the present case, if we want to fix the particle density to its critical 
value $\zeta_c$.  Since the height is
a discrete variable, some local variations in density and activity are
inevitable.  All the results described up to now were obtained using the
random deposition (RD) initial preparation described in Sec. II.  In this 
subsection we report on studies using initial configurations in which the 
$N = \zeta L$ particles are distributed so as to reduce density fluctuations.  
We again employ the sequence $L=2^n \cdot 1253$, $N = 2^n \cdot 1189$ 
($n=0,...,4$) corresponding to $\zeta = 0.94892$.

One way of suppressing density fluctuations is via {\it restricted random
deposition} (RRD): we divide the system into blocks, each consisting of 25 
(or in some cases 26)  sites, and depositing at random, within each block,
a fixed number (24 or 25) of particles.  In this way we generate a height 
distribution very similar to that of standard RD, but with virtually no 
density fluctuations on scales $> 50$ sites.  
Our conclusions from RRD simulations are as follows.  
First, the stationary activity density is identical, to within
uncertainty, to that found previously.  Thus we may continue to use 
$\beta/\nu_{\perp} = 0.239$ in the scaling analysis.  
Figure 10, a scaling plot of the activity
density (as in Fig.~8), shows a reasonable data
collapse at long times, whereas the initial phase does not collapse.  (It {\it does}
collapse in an unrescaled plot, showing that the early stage is a size-independent,
presumably purely local process.)   Once again $z$ has been chosen to
optimize the collapse at long times, which in this case gives $z=1.47(6)$.
The final approach to the stationary state
appears to follow a power law, $\rho_a \sim t^{-\theta}$, with 
$\theta = 0.16(1)$.
Analysis of the interface representation yields $z=1.57(3)$,
$\alpha = 1.13(2)$ and $\beta_W = 0.77(3)$, consistent
with the scaling relation $\beta_W = \alpha/z$.  
(The growth of the survival time with $L$ suggests yet another
value, $z \simeq 1.3$.)
Thus for RRD, all of the exponents describing transient properties are different
from those found in RD studies.  The exponent $z$ has decreased to a value
consistent with stationary relaxation, perhaps due to suppression of
slowly relaxing, long-wavelength density variations.  The sum $\theta + \beta_W$,
however, is now somewhat smaller than unity.

A second kind of initial configuration with reduced density fluctuations may be termed
{\it maximum activity} (MA).  In this case all sites are
either empty or have height 2 (except for a single site, in case $N$
is odd), and the spacing between sites with $z=2$ is fixed at 1 or 2 sites,
in a regular pattern, so the average density is constant on a scale of 50 or
so sites.   While the density is again uniform on large scales, the number of
active sites is as large as possible (as in usual initial-decay studies), 
and consequently the height distribution
is quite different from that in the RD or RRD configurations.  
For MA IC's we find $\alpha = 1.12(4)$, $\beta_W = 0.78(1)$, and
$\theta = 0.168(6)$.  From the collapse of $\rho_a (t)$ and of $W^2 (t)$,
we obtain, respectively, $z=1.47(4)$ and $z=1.49(2)$, while the survival time
data again yield $z \simeq 1.3$.  Thus our results for the two kinds of
initial conditions, RRD and MA, with suppressed density fluctuations are
in good agreement.  This suggests that the precise form of the initial
height distribution does not affect asymptotic scaling behavior.

A general conclusion from these studies is that suppressing density fluctuations
reduces the apparent value of the exponent $z$, suggesting that the unusually
slow relaxation observed using random deposition initial conditions is due to
long-wavelength density variations.  The exponents $\alpha$ and $\beta_W$
decrease as well, indicating that such density variations also provide a source of
roughness for the interface dynamics.  
Further studies are required to determine whether the interface-growth exponents
reported here are typical of an entire class of initial conditions (i.e., with
reduced density fluctuations), or if their values may depend on other
details of the initial preparation.  Until such studies are performed, we 
cannot be sure that the exponents describing roughness and initial relaxation
are well defined, describing asymptotic evolution, or merely represent 
apparent power laws over some limited range of times and system sizes.
In none of the cases studied do we observe
the simple initial relaxation of $\rho_a$ typical of the contact process.  
This may be due to the impossibility of preparing a completely uniform, 
high-activity initial state, and/or to the conservation of particles.

\subsection{Ergodicity} 
 
In a recent study we found strong nonergodic 
effects in the BTW sandpile \cite{fes2}. 
Given the randomness in the Manna sandpile rules, we do not
expect such effects here.
To test this hypothesis we plot (see Fig.~11) 
the active-site density $\rho_a$ versus time in four individual 
trials in a system of 1000 sites near $\zeta_c$.  
The curves for $\rho_a$ in the 
various trials are completely intermixed; their ranges of 
fluctuation about a common mean 
are roughly equal. This is in complete contrast to the BTW case, 
where a similar plot shows distinct mean values, and
very different ranges of variation of $\rho_a$, in
each trial \cite{fes2}.   
Thus the Manna sandpile appears 
to be ergodic, in the sense that a time-average of $\rho_a$ 
in a single realization 
is independent of the initial condition (after a short transient), 
and yields the 
same value as the stationary average of $\rho_a$ over many trials.  
Nonergodicity in the BTW 
model has been associated with ``hidden'' conservation laws, 
i.e., a memory of the initial 
configuration that persists over the entire trial.  
The Manna sandpile seems to be 
free of such conservation laws: while we can not rigorously exclude their 
presence, their effect must in any event be much weaker than 
in the BTW case.  This is of course consistent with 
the fact that the BTW dynamics 
is deterministic, while the redistribution of particles 
is stochastic in the present case.  (As a simple example of this, recall
that a nearest-neighbor pair with both sites empty ($z=0$) is forbidden
in the stationary state of the BTW sandpile, whereas no such restriction
exists for stochastic toppling rules.)
 
\section{Discussion} 
 
We studied the scaling behavior of a one-dimensional 
fixed-energy sandpile with 
the same local dynamics as the Manna model.   
The model exhibits a continuous phase transition between an absorbing  
state and an active one at a critical particle density 
$\zeta_c = 0.94885(7)$.  In the interface 
(integrated activity) picture this translates into a depinning 
transition with a fixed force (fixed-energy model) or a force 
that is ramped up at an infinitesimal rate (driven case) \cite{lau}. 
 
The phase transition in the one-dimensional stochastic sandpile  
is characterized by the critical exponents 
$\beta=0.42(2)$ and $\nu_{\perp} = 1.81(1)$, 
which differ significantly from 
those associated with directed percolation ($\beta=0.2765$,  
$\nu_{\perp} = 1.0968$) 
and linear-interface depinning ($\beta=0.25(3)$, 
$\nu_{\perp} \simeq 1.3$).
While absorbing-state phase transitions are expected to fall generically 
in the directed percolation universality class \cite{janssen,gr82}, it is 
reasonable to exempt the Manna model from this rule, due to 
local conservation of particles; this conservation law is expected to alter
the universality class.
In fact,  studies of various models with the 
same local conservation law as the Manna sandpile, 
in dimensions $d >1$, 
indicate a new, common 
universality class for models sharing this feature \cite{fes2,alepp,pv}. 
 
Studying the interface representation of the model, 
we obtain the roughness exponent 
$\alpha = 1.48(2)$ and
growth exponent $\beta_W = 0.86(2)$, which should be 
compared with $\alpha = 1.33(1)$, $\beta_W = 0.839(1)$ for DP and 
$\alpha = 1.25(1)$, $\beta_W = 0.88(2)$ for LIM. 
Study of the height-height correlation function yields the local roughness 
exponent $\alpha_{loc} = 0.59(3)$; 
the corresponding DP value is $0.63(3)$ \cite{rdmam}. 
Changing the initial condition to suppress long-wavelength density fluctuations,
we obtain $\alpha = 1.13(2)$ and
$\beta_W = 0.77(2)$, which again differ from the values associated with DP and LIM.
Comparing these results with 
the apparent agreement of the Manna exponents with those of the LIM in 
two dimensions implies that either the rough numerical equivalence is 
fortuitous, or that the noise $\tau$ in the interface equation has 
a fundamentally different structure depending on the dimension. 
It is worth remarking that the measured roughness exponent
is rather close to $\alpha=3/2$, which is the value one expects
if only the columnar component of the noise 
is relevant \cite{pp91}.
The other exponents, however, seem to be far from the columnar
disorder universality class 
(i.e., $\nu_\perp=2$, $z=2$, $\beta=1$, $\beta_w=3/4$) \cite{pp91}.

In linear interface models, translational invariance of the 
noise can be used to derive the scaling relation 
$(2-\alpha)\nu_\perp=1$
\cite{kardar}. 
This relation does not appear to be verified by our numerical results. 
Notice that in the 
context of interface depinning $z$ can be linked to the other
exponents by a scaling relation \cite{kardar}
which reads $z=\beta/\nu_\perp + \alpha$. Inserting the values
of $\beta$, $\nu_\perp$ and $\alpha$ measured in simulations,
we obtain $z=1.7$, which is consistent with the exponent
obtained analyzing the scaling of the width, for RD initial conditions.
Our results for initial conditions with reduced fluctuations do not,
however, follow this scaling relation. Understanding this in
the interface picture by considering the height-dependent
quenched noise term is a challenge.

Our results for the dynamic exponent $z$ are conflicting.  
We have studied various definitions of the relaxation or
correlation time: that for the decay of the activity autocorrelation
in the stationary state (yielding $z=1.49(3)$), one associated with
the survival time ($z=1.42(1)$ for RD initial conditions, $z \simeq 1.3$
for RRD and MA IC's), and those associated with the surface roughness
and the decay of activity ($z = 1.70(5)$ for RD IC's, $z = 1.52(5)$
for initial configurations with suppressed density fluctuations).
At a simple absorbing-state phase transition, all of these relaxation
times are of course governed by the same exponent.
Whether the unusually large $z$ associated with time-dependent behavior
using random deposition derives, as we have suggested, from slow
relaxation of long-wavelength density fluctuations in the initial
configuration, is a subject for further investigation.  
Our studies of modified initial
states support this notion, but to reach a definitive conclusion
other classes of initial conditions will need to be examined. 
Indeed, we have
noted that, given their apparent dependence on the initial state,
it is not clear whether the exponents $z$, $\alpha$, $\beta_W$ and
$\theta$ are well defined, i.e., whether the asymptotic behavior
follows simple power laws.
 
We note that the present model does not exhibit the strong 
nonergodic effects observed in the fixed-energy version of the BTW sandpile. 
The relaxation of the mean height $\langle z_i (t) \rangle$ from its initial
value to the average, $\zeta$, follows a power law with an exponent $ \approx 1/2$.
We find good evidence for finite-size scaling, in contrast with 
most driven sandpiles \cite{kad89,stella,drossel}. 
In summary, we have identified a one-dimensional sandpile model 
that exhibits an absorbing-state phase transition as 
the relevant temperature-like 
parameter (the energy density) is varied.  It appears to be the 
``minimal model'' for absorbing-state phase transitions 
belonging to a new universality 
class associated with a conserved density. 
Preliminary studies indicate that the 
driven version of the model exhibits scale-invariant avalanche 
statistics \cite{bjp,rdpp}. 
We may therefore hope that analysis of the driven model, and of spreading  
of activity at the critical point of the fixed-energy system, 
will permit us to  
establish detailed connections 
between scale invariance under driving and the underlying absorbing-state  
phase transition. 

Let us finally stress that while in higher dimensions the linear
interface universality class and that of systems with absorbing states
in the presence of a conserved static field seem to coincide, the 
results of this paper show that this equivalence appears to be
broken in $d=1$. It will be interesting 
to study other one-dimensional
systems with absorbing states and an order parameter coupled
to a static conserved field \cite{AIP,alepp,pv}
in order to compare the critical exponents and anomalies with
those reported in this paper.
\vspace{1em}

{\bf Acknowledgments}
\vspace{.5em}

We thank R. Pastor-Satorras and F. van Wijland for helpful
comments and discussions.
RD thanks CNPq for financial support and CAPES and FAPEMG for support
of computer facilities.
We acknowledge partial support
from the European Network contract ERBFMRXCT980183.
MAM acknowledges support from the
Spanish DGESIC project PB97-0842, and
Junta de Andaluc{\'\i}a project FQM-165.

\newpage 
\noindent FIGURE CAPTIONS 
\vspace{1em} 
 
\noindent FIG. 1.  
Stationary active-site density versus energy density $\zeta$. 
The points represent extrapolations of data for  
$L = 100 $ - 5000 to the $L \to \infty $ limit. 
\vspace{1em} 
 
\noindent FIG. 2.  
Stationary active-site density versus system size. 
From bottom to top, $\zeta$ = 0.948, 0.94857, 0.94864, 0.94874, 0.9488, 
0.94892, 0.949, and 0.95.  The inset shows the curvature $b$ of the log-log plot  
as a function of $\zeta$, for $L \geq 1000$.  The straight line is a least-squares 
linear fit. 
\vspace{1em} 
 
\noindent FIG. 3.   
Inset: stationary autocorrelation function $C(t)$  versus $t$ 
at $\zeta = 0.9488$, for (left to right) $L = 625$, 1250, 
2500, 5000, and $10^4$.  In the main graph the data are plotted versus 
a rescaled time $t^*$ defined in the text. 
\vspace{1em} 
 
\noindent FIG. 4.  
Scaling plot of the active-site density versus 
$\Delta \equiv \zeta - \zeta_c$.  Symbols: $+$: $L=100$; 
$\bullet$: $L=200$; $\times$: $L=500$; 
$\circ$: $L=1000$; $\Box$: $L=2000$; $\Diamond$: $L=5000$. 
\vspace{1em} 
 
\noindent FIG. 5.  
Scaling plot of $\tilde{W}^2 \equiv W^2/L^{2 \alpha}$ versus 
$\tilde{t} \equiv t/L^z$, for $\zeta = 0.9489$, using $\alpha = 1.41$, and 
$z = 1.645(13)$.  System sizes (top to bottom) $L= 1253$, 2506, 5012, 10024,  
and 20048. 
\vspace{1em} 
 
\noindent FIG. 6.  
Scaling plot of saturation width versus lifetime in individual trials, 
$\zeta = 0.9490$, $L=400$-6400. 
\vspace{1em} 
 
\noindent FIG. 7.  
Saturation height versus lifetime in individual trials, 
$\zeta = 0.9490$, $L=400$-6400. 
\vspace{1em} 
 
\noindent FIG. 8.  
Scaled active-site density $\rho^* = L^{\beta/\nu_\perp} \rho_a $
versus scaled time $t^* = t/L^z$ for $\zeta = 0.9489$, 
system sizes $L=1253$,..., 20048 as indicated. 
\vspace{1em} 
 
\noindent FIG. 9.  
Mean height $\langle z(t)|z(0)\rangle$ of sites with 
initial height $z(0)$, in at system of 1400 sites at $\zeta_c$, averaged 
over 2000 trials.  From bottom to top: $z(0)$ = 0, 1, 2, 3, and 4. 
The inset is a plot of $\ln |\langle z(t)|z(0)\rangle - \zeta| $ versus $\ln t$. 
\vspace{1em} 

\noindent FIG. 10.  
Scaled active-site density versus scaled time for $\zeta = 0.9489$ as in
Fig.~8, but for RRD initial conditions.
\vspace{1em} 

\noindent FIG. 11.  
Active-site density $\rho_a(t)$ in four different trials at $\zeta = 0.95$ 
in a system of 1000 sites. 
\vspace{1em} 
 

\begin{references} 
 
\bibitem{btw} 
P. Bak, C. Tang, and K. Wiesenfeld, 
Phys. Rev. Lett. {\bf 59}, 381 (1987); 
Phys. Rev. A {\bf 38}, 364 (1988). 
 
\bibitem{ggrin}    
G. Grinstein, in  
{\it Scale Invariance, Interfaces and Nonequilibrium Dynamics},  
{\it NATO Advanced Study Institute, Series B: Physics}, 
vol. 344, A. McKane et al., Eds.  
(Plenum, New York, 1995). 
 
\bibitem{bjp} 
R. Dickman, M. A. Mu\~noz, A. Vespignani, and S. Zapperi, 
Braz. J. Phys. {\bf 30}, 27 (2000). 
    e-print: cond-mat/9910454. 

\bibitem{AIP}
M. A. Mu\~noz, R. Dickman, R. Pastor-Satorras,
A. Vespignani, and S. Zapperi, ``Sandpiles and absorbing state
phase transitions: recent results and open problems,"
to appear in {\it Procedings of the $6th$ Granada seminar on
computational physics}, Ed. J. Marro and P. L. Garrido. (American
Institute of Physics, 2001). e-print: cond-mat/0011447.    
 
\bibitem{tb88} 
C. Tang and P. Bak,  
Phys. Rev. Lett. {\bf 60}, 2347 (1988). 
 
\bibitem{vz} 
A. Vespignani and S. Zapperi, 
Phys. Rev. Lett. {\bf 78}, 4793 (1997); 
Phys. Rev. E {\bf 57}, 6345  (1998). 
 
\bibitem{dvz} 
R. Dickman, A. Vespignani, and S. Zapperi, 
Phys. Rev. E {\bf 57}, 5095 (1998). 
 
\bibitem{vdmz} 
A. Vespignani, R. Dickman, M. A. Mu\~noz, and Stefano Zapperi,  
Phys. Rev. Lett. {\bf 81}, 5676 (1998). 
 
\bibitem{mdvz} 
M. A. Mu\~noz,  R. Dickman, A. Vespignani, and Stefano Zapperi,  
Phys. Rev. E {\bf 59}, 6175 (1999).  
 
\bibitem{chessa}  
A. Chessa, E. Marinari and A. Vespignani, 
Phys. Rev. Lett. {\bf 80}, 4217 (1998). 
 
\bibitem{carlson} 
A. Montakhab and J. M. Carlson,  
Phys. Rev. E {\bf 58}, 5608 (1998). 

\bibitem{fes2} 
A. Vespignani, R. Dickman, M. A. Mu\~noz, and S. Zapperi, 
Phys. Rev. E {\bf 62}, 4564 (2000). 
  
\bibitem{dhar99}  
D. Dhar, Physica A {\bf 263}, 4 (1999), and references therein. 
 
\bibitem{priez}V. B. Priezzhev, J. Stat. Phys. {\bf 74}, 955 (1994); 
E. V. Ivashkevich, J. Phys. A {\bf 27}, 3643 (1994); 
E. V. Ivashkevich, D. V. Ktitarev and V. B. Priezzhev, 
Physica A {\bf 209}, 347 (1994).  
 
\bibitem{manna} 
S. S. Manna, 
J. Phys. A {\bf 24}, L363 (1991). 
 
\bibitem{manna2} 
S. S. Manna,   
J. Stat. Phys. {\bf 59}, 509 (1990). 

\bibitem{midd}  
O. Narayan and A. A. Middleton,  
Phys. Rev. B {\bf 49} 244 (1994). 
 
\bibitem{pacz}  
M. Paczuski and S. Boettcher,   
Phys. Rev. Lett. {\bf 77}, 111 (1996). 
 
\bibitem{lau} 
K. B. Lauritsen and M. Alava,  
cond-mat/9903346. 

\bibitem{ala} 
M. Alava and K. B. Lauritsen, 
cond-mat/0002406, to be published in Europhys. Lett.

\bibitem{alepp} 
M. Rossi, R. Pastor-Satorras, and A. Vespignani, 
Phys. Rev. Lett. {\bf 85}, 1803 (2000).

\bibitem{pv}
R. Pastor-Satorras and A. Vespignani, 
Phys. Rev. E {\bf 62}, R5875 (2000).
 
\bibitem{fss}  
M. E. Fisher, in {\it Proceedings  
of the International Summer School `Enrico 
Fermi', Course LI}, (Academic Press, New York, 1971); 
M. E. Fisher and M. N. Barber,  
Phys. Rev. Lett. {\bf 28}, 1516 (1972);
{\it Finite Size Scaling}, Ed. J. Cardy, North-Holland (Amsterdam),
1988.

\bibitem{barabasi} 
    A. -L. Barab\'asi and H. E. Stanley, 
    {\it Fractal Concepts in Surface Growth}, 
    (Cambridge University Press, Cambridge, 1995). 
 
\bibitem{hhz} 
T. Halpin-Healy and Y.-C. Zhang, 
Phys. Rep. {\bf 254}, 215 (1995). 
 
\bibitem{Lesch}  
H. Leschhorn, Physica A {\bf 195}, 324 (1993).

\bibitem{Lopez_97} 
J.M.\ L\'opez, M.A.\ Rodr\'iguez, and R. Cuerno, 
Phys.\ Rev.\ {\bf E 56}, 3993 (1997); 
	J. M. L\'opez,
	Phys. Rev. Lett. {\bf 83}, 4594 (1999);
 
\bibitem{FV}  
F. Family and T. Vicsek, J. Phys. A  {\bf 18}, L75 (1985). 
 
\bibitem{marro} 
 J. Marro and R. Dickman,  
 {\em Nonequilibrium Phase Transitions in Lattice Models}  
 (Cambridge University Press, Cambridge, 1999). 
 
\bibitem{rdmam} 
R. Dickman and M. A. Mu\~noz, Phys. Rev. E {\bf 62},
7632 (2000). 
 
\bibitem{sv99}
J. Schmittbhul and J.-P. Villotte,
Physica A, {\bf 270}, 42 (1999).
\bibitem{cgl00}
P. Chauve. T. Giamarchi and P. Le Doussal,
Phys. Rev. B {\bf 62}, 6241 (2000).
 
\bibitem{krug} 
J. Krug, 
Adv. Phys. {\bf 46}, 139 (1997). 
 
 
\bibitem{kad89} 
L. P. Kadanoff, S. R. Nagel, L. Wu and S. Zhou,  
Phys. Rev. A {\bf 39}, 6524 (1989). 
 
\bibitem{stella} 
M. De Menech, A. L. Stella, and C. Tebaldi,  
    Phys. Rev. E {\bf 58}, R2677 (1998); 
C. Tebaldi, M. De Menech, and A. L. Stella, 
Phys. Rev. Lett. {\bf 83}, 3952 (1999). 
 
\bibitem{drossel} 
B. Drossel, 
    Phys. Rev. E {\bf 61}, R2168 (2000). 
 
\bibitem{janssen} 
H. K. Janssen, 
Z. Phys. {\bf 42}, 141 (1981); 
{\it ibid.} {\bf 58}, 311 (1985). 
 
\bibitem{gr82} 
P. Grassberger, 
Z. Physik. B {\bf 47}, 465 (1982). 
 
\bibitem{pp91}
G. Parisi and L. Pietronero,
Europhys. Lett. {\bf 16},  321 (1991);
Physica A {\bf 179}, 16 (1991).

\bibitem{kardar}
        M. Kardar,
        Phys. Rep. {\bf 301}, 85 (1998). 

\bibitem{rdpp} 
        R. Dickman, unpublished. 
 
\end{references}
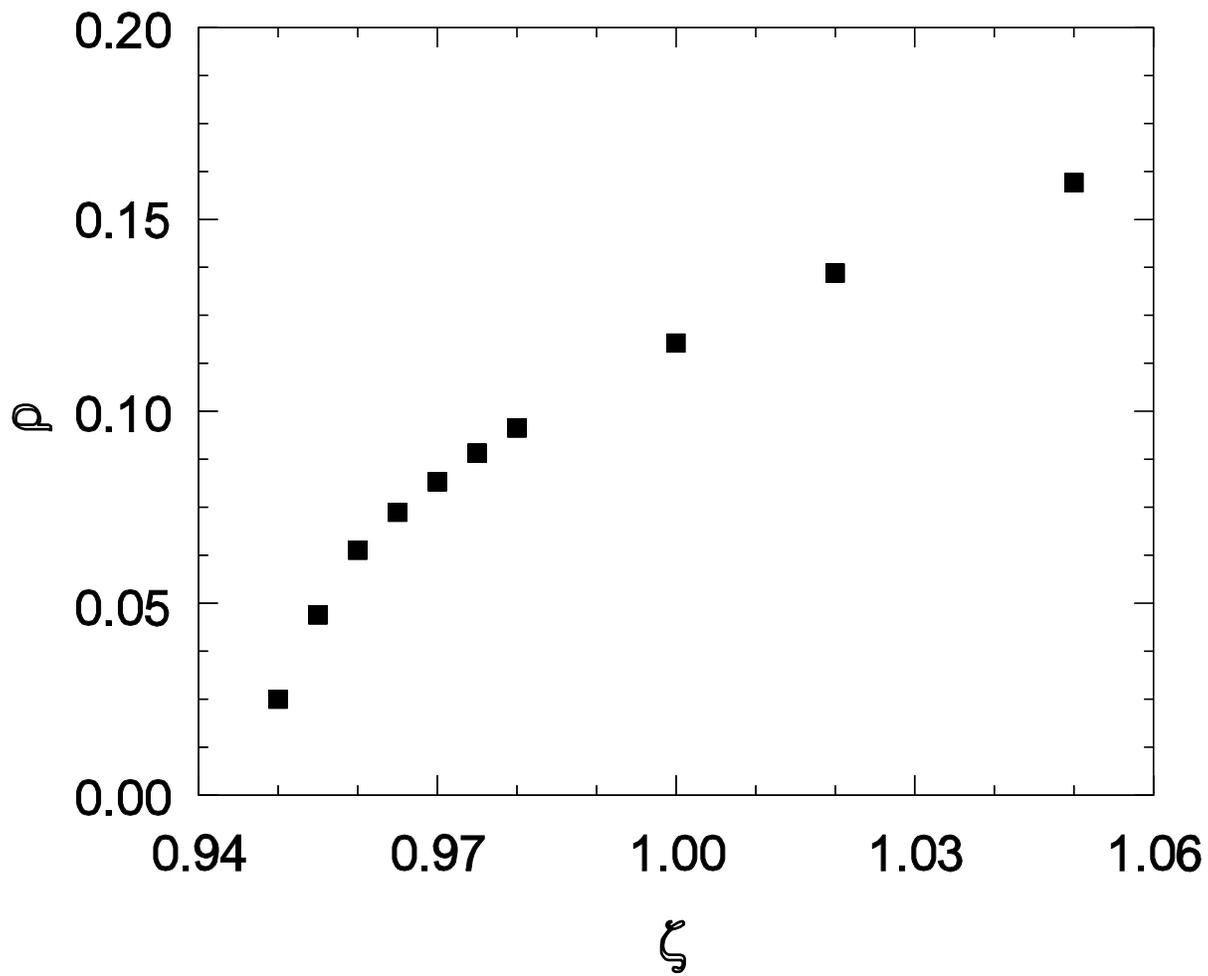
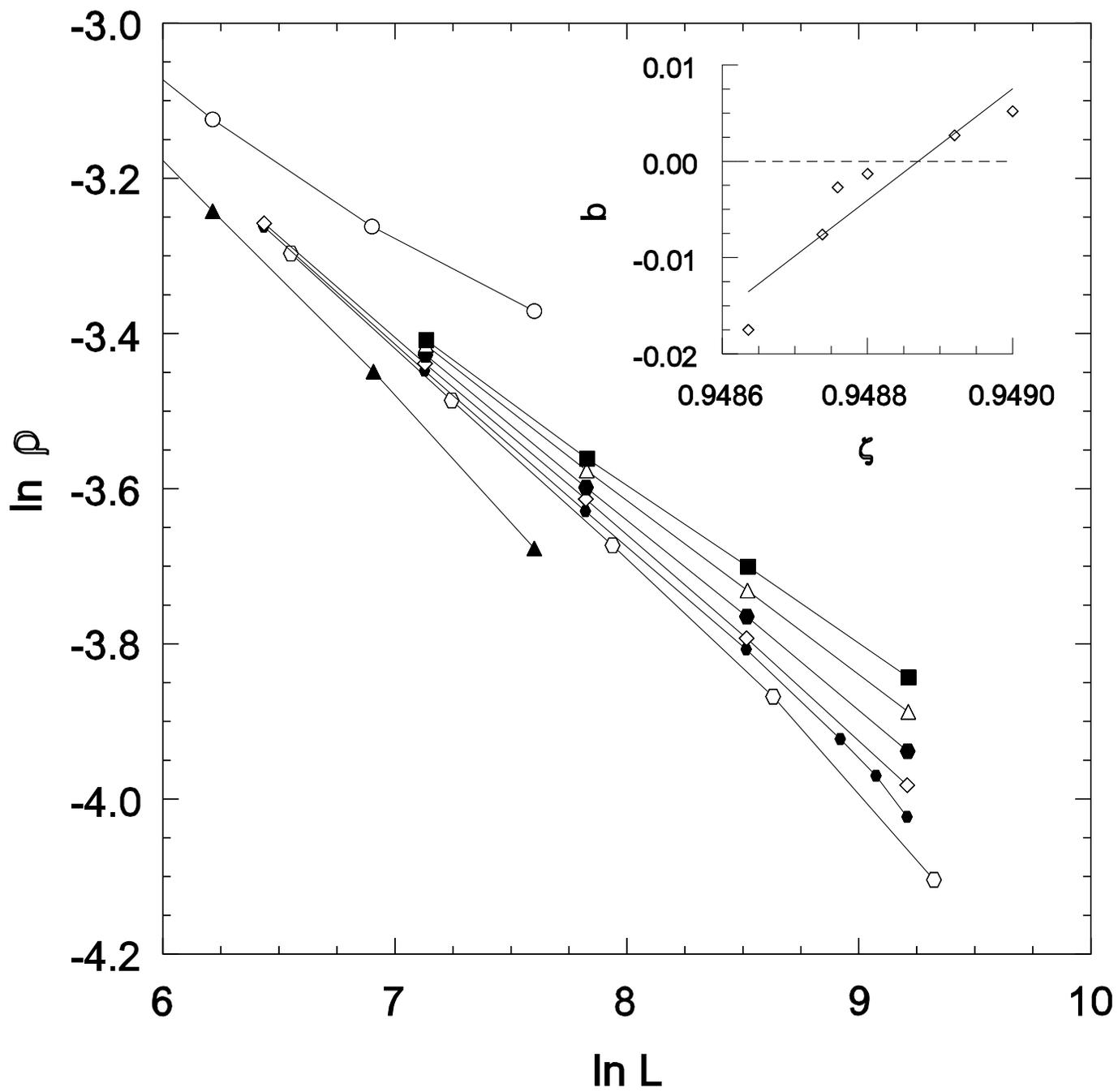
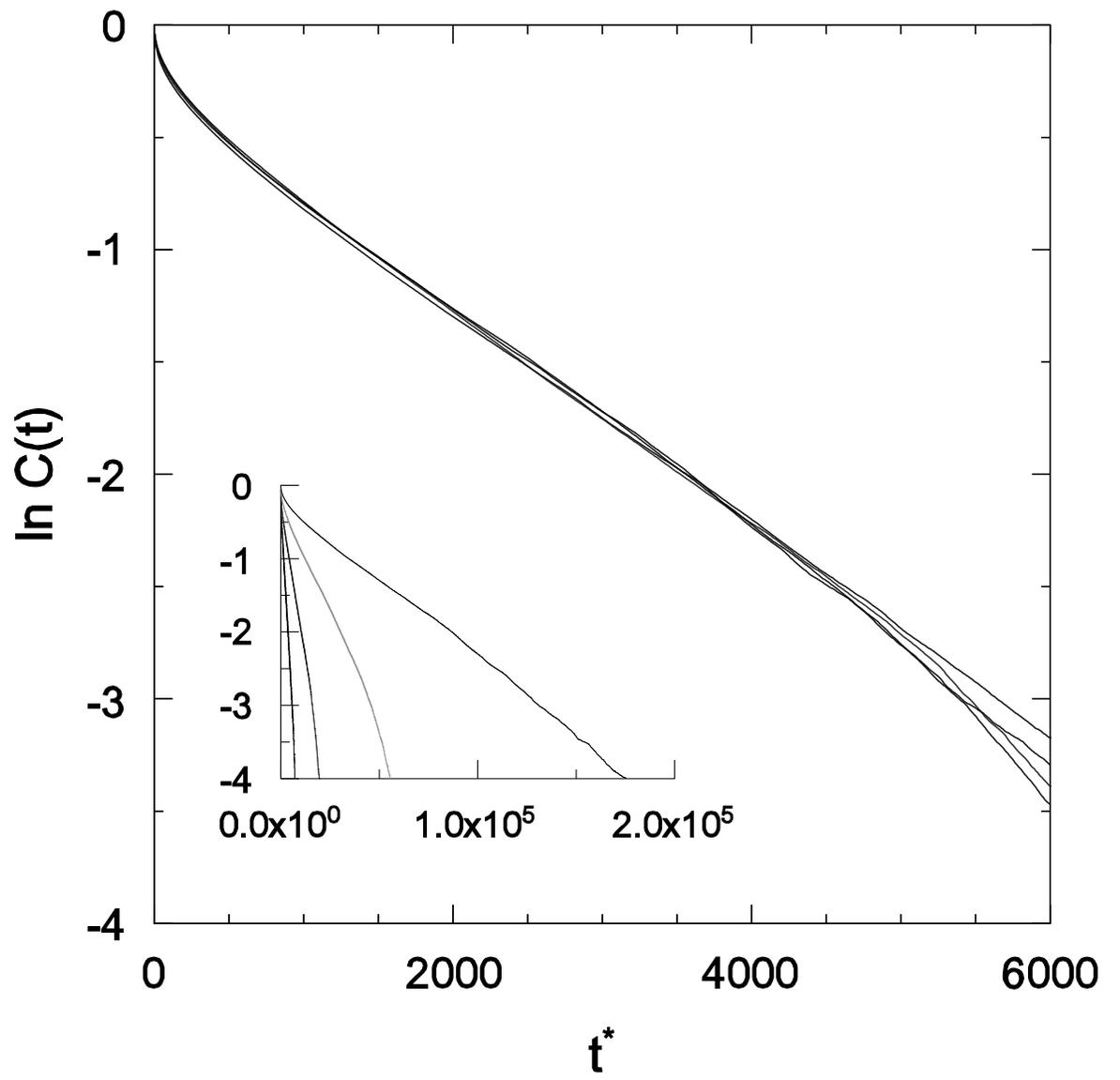
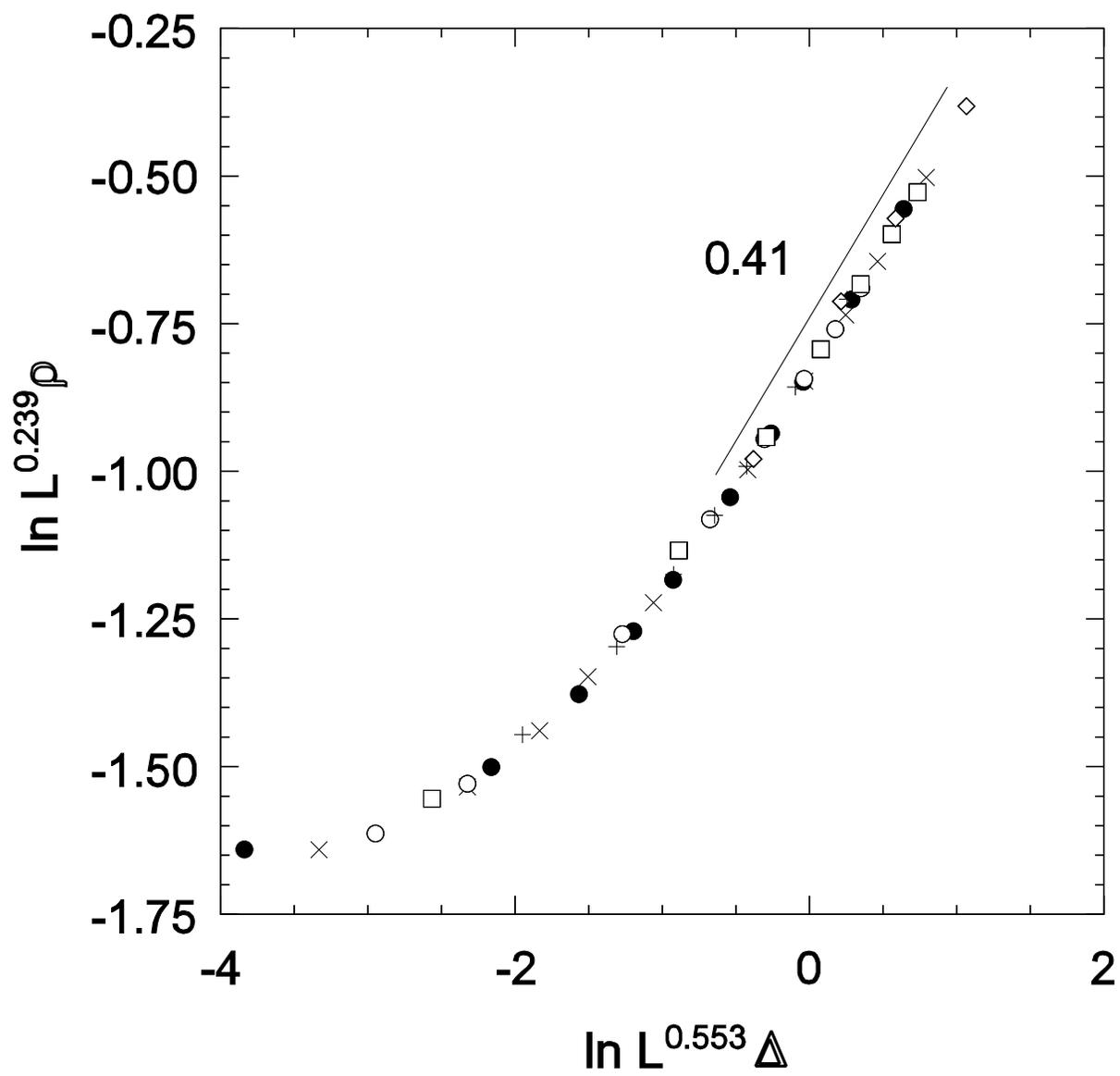
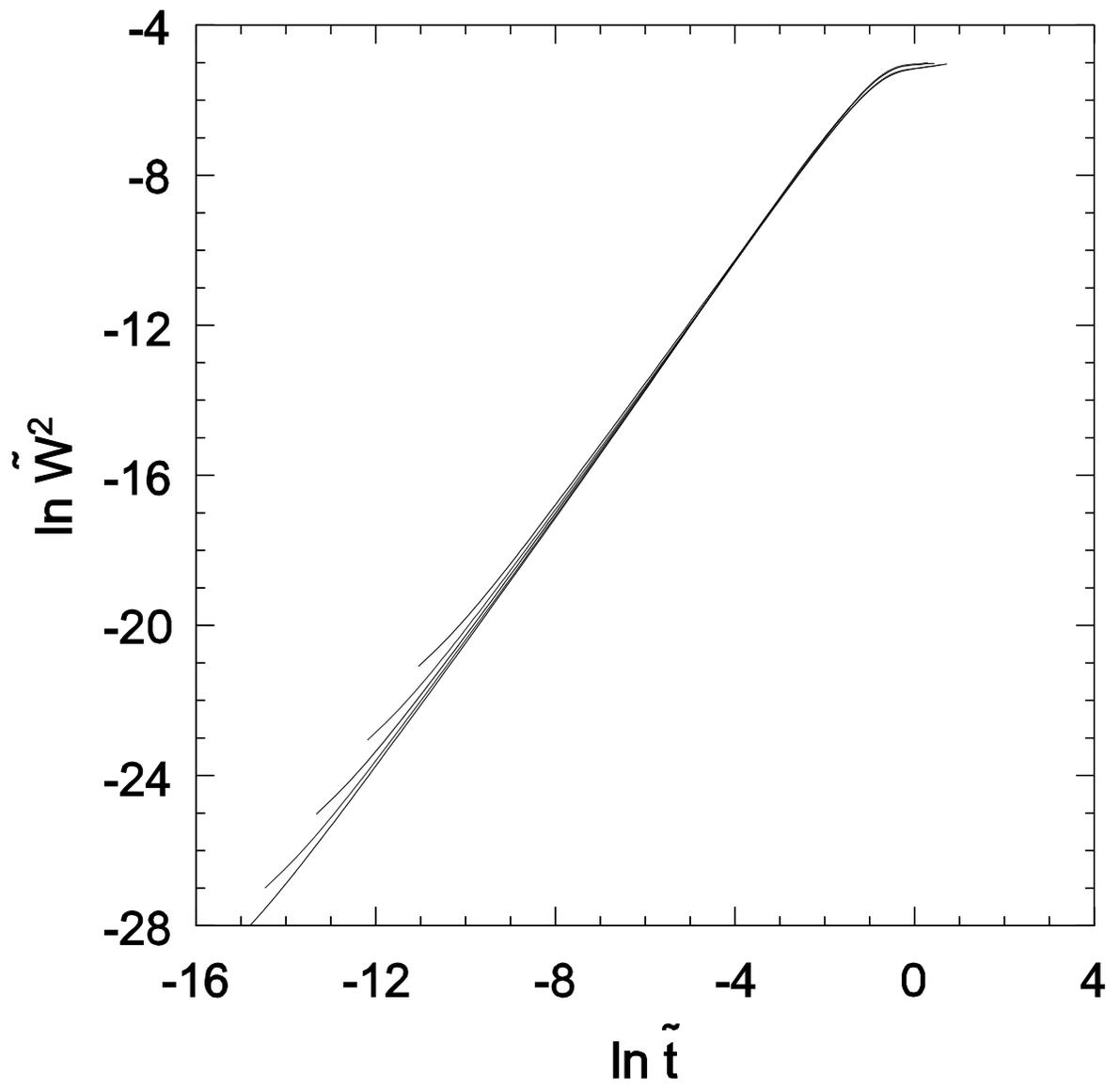
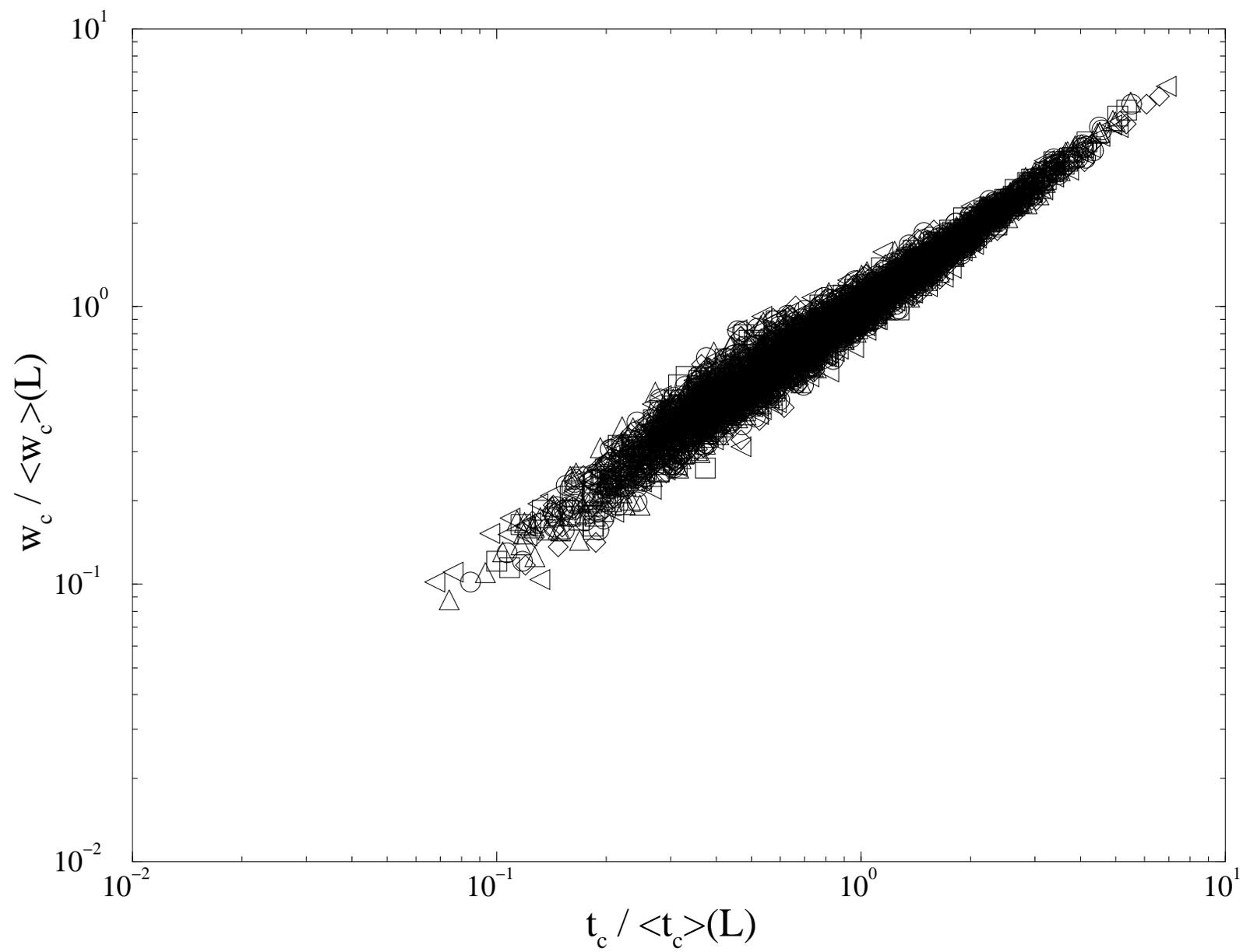
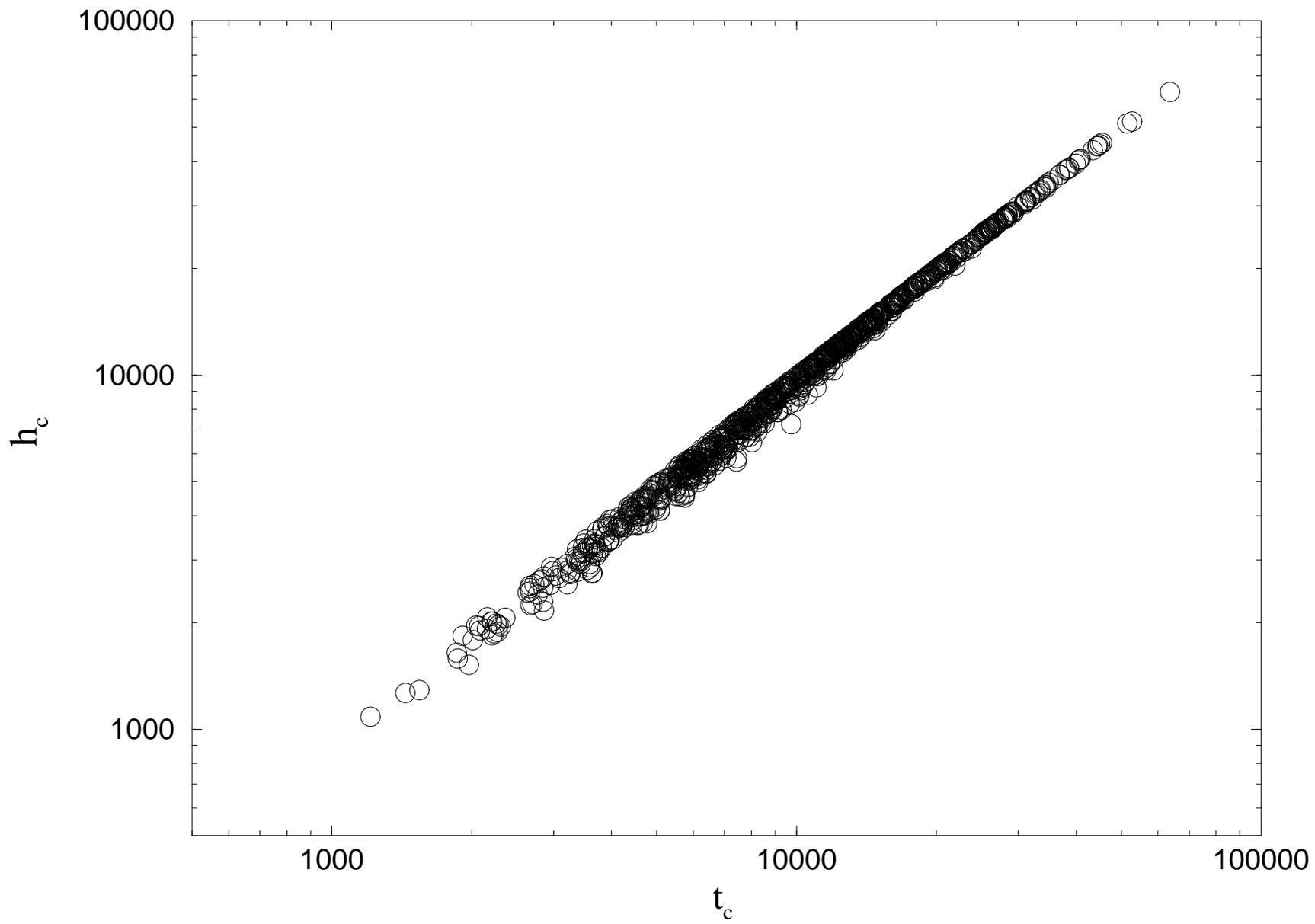
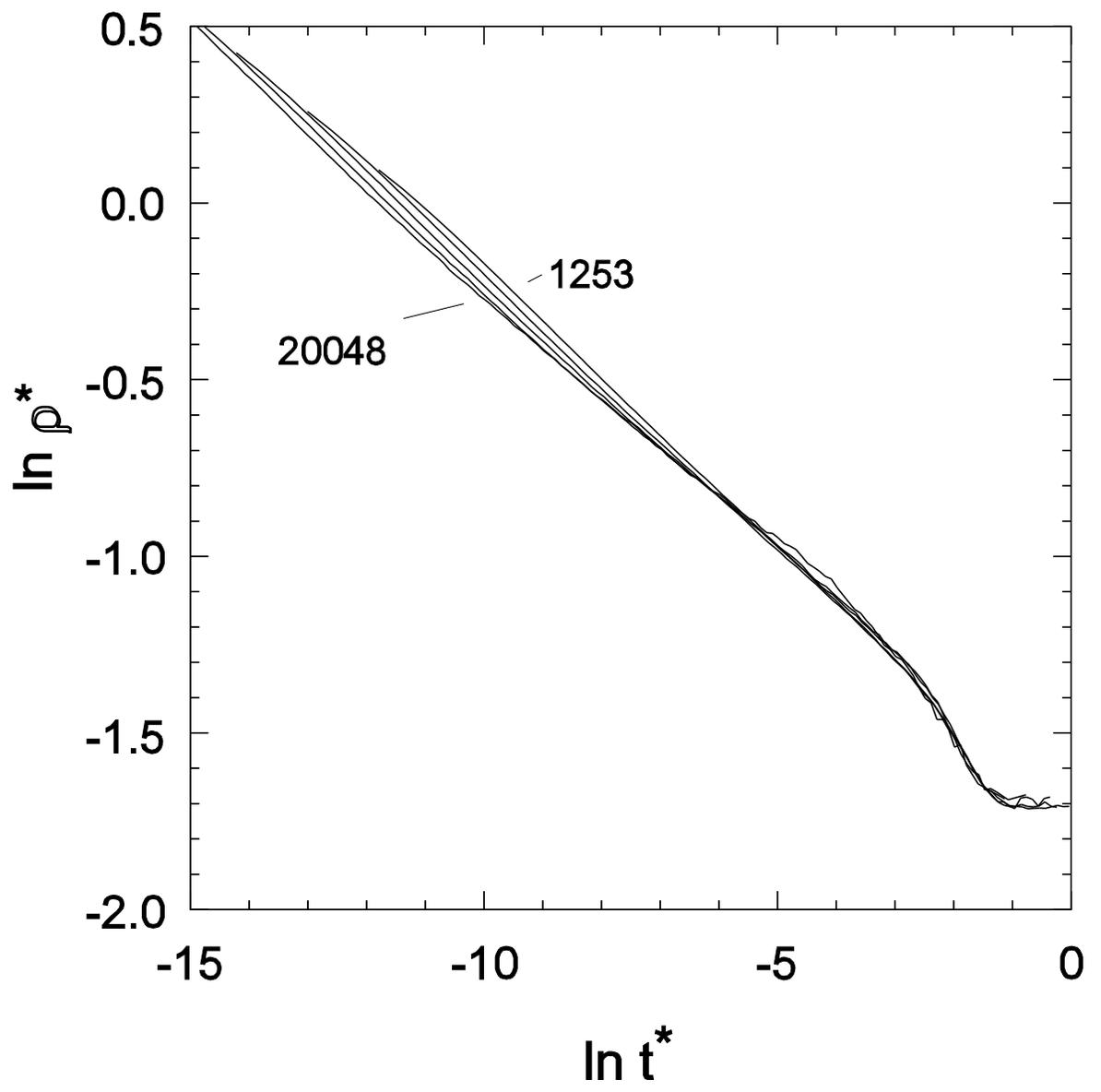
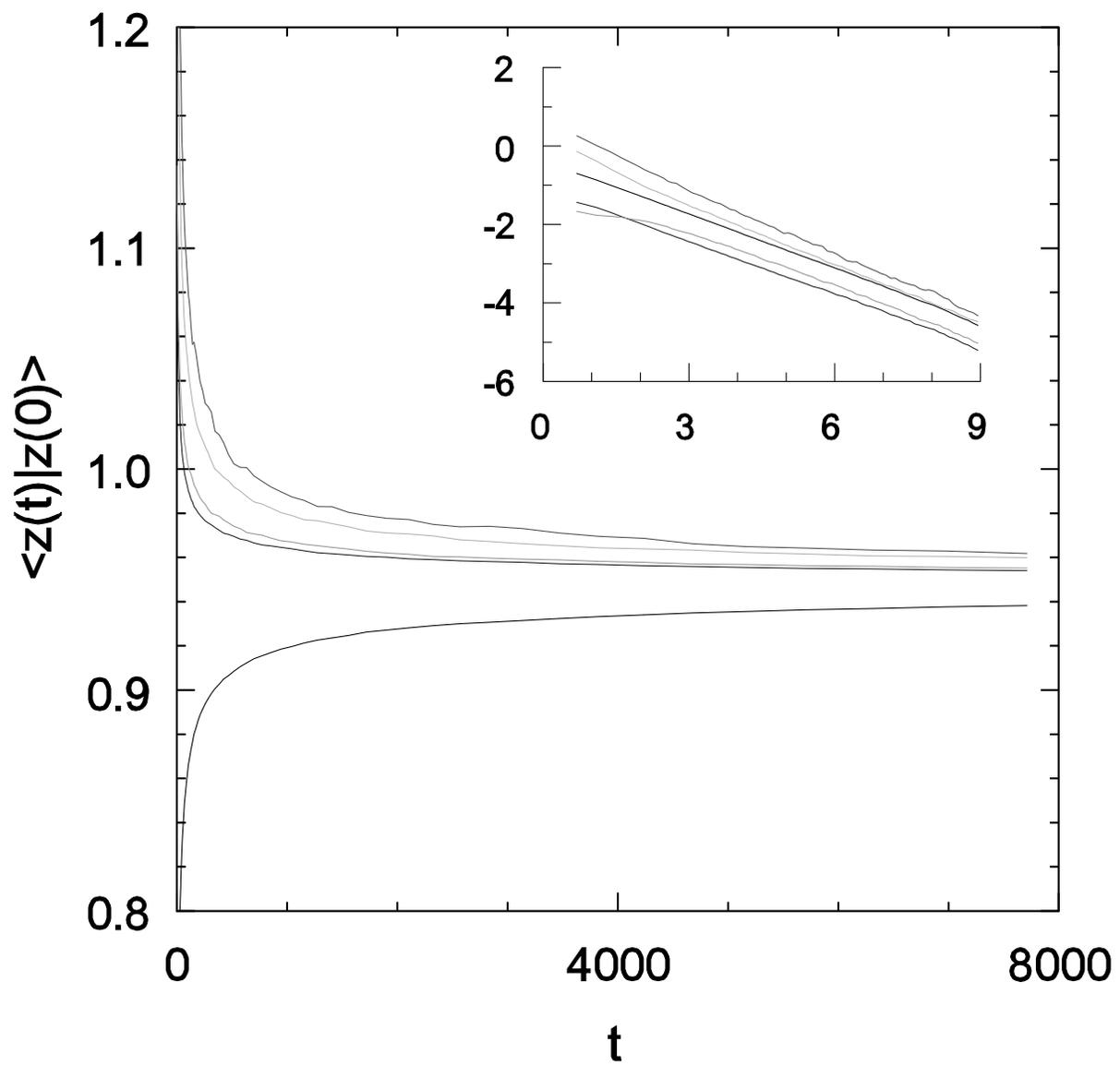
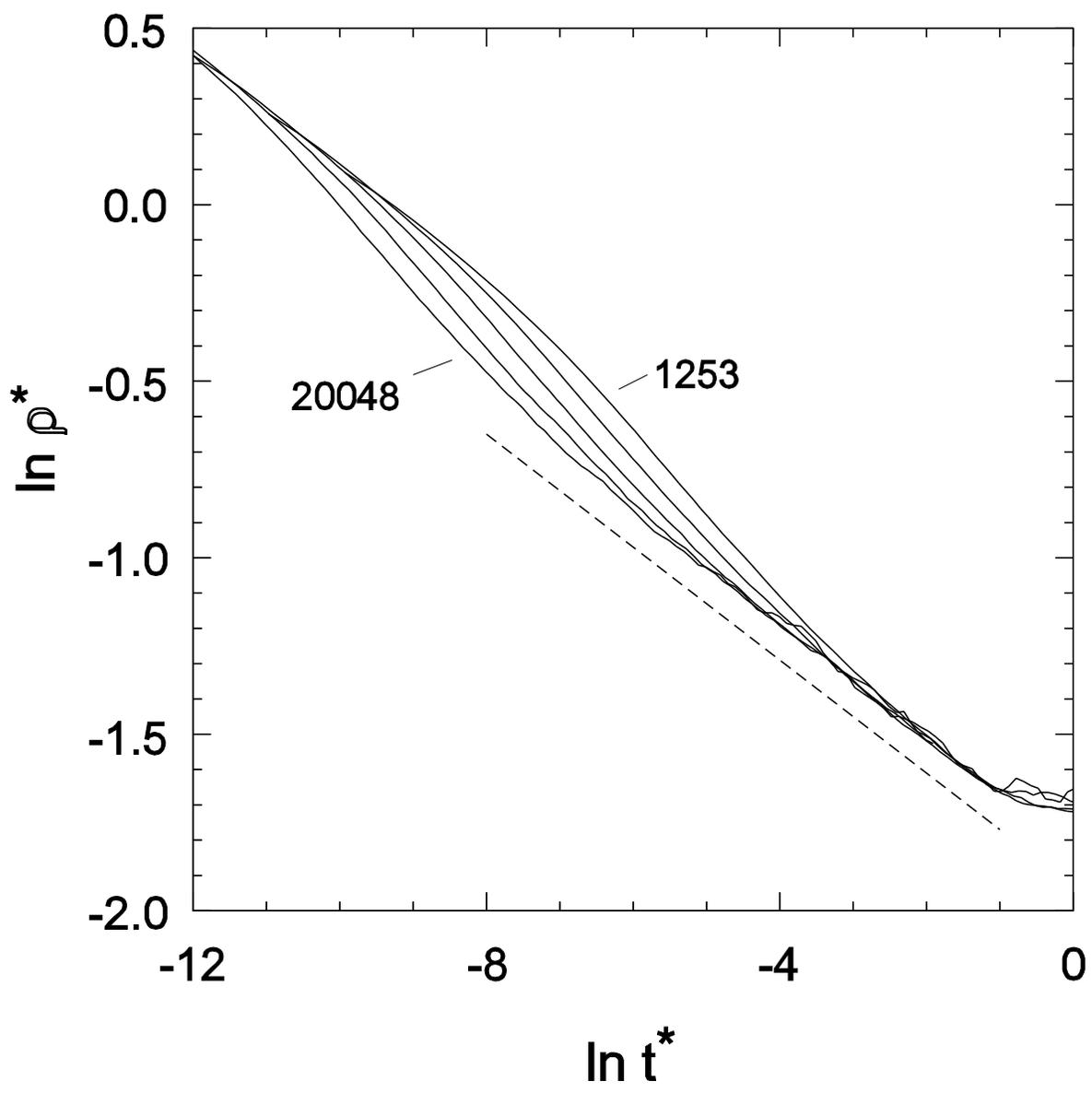
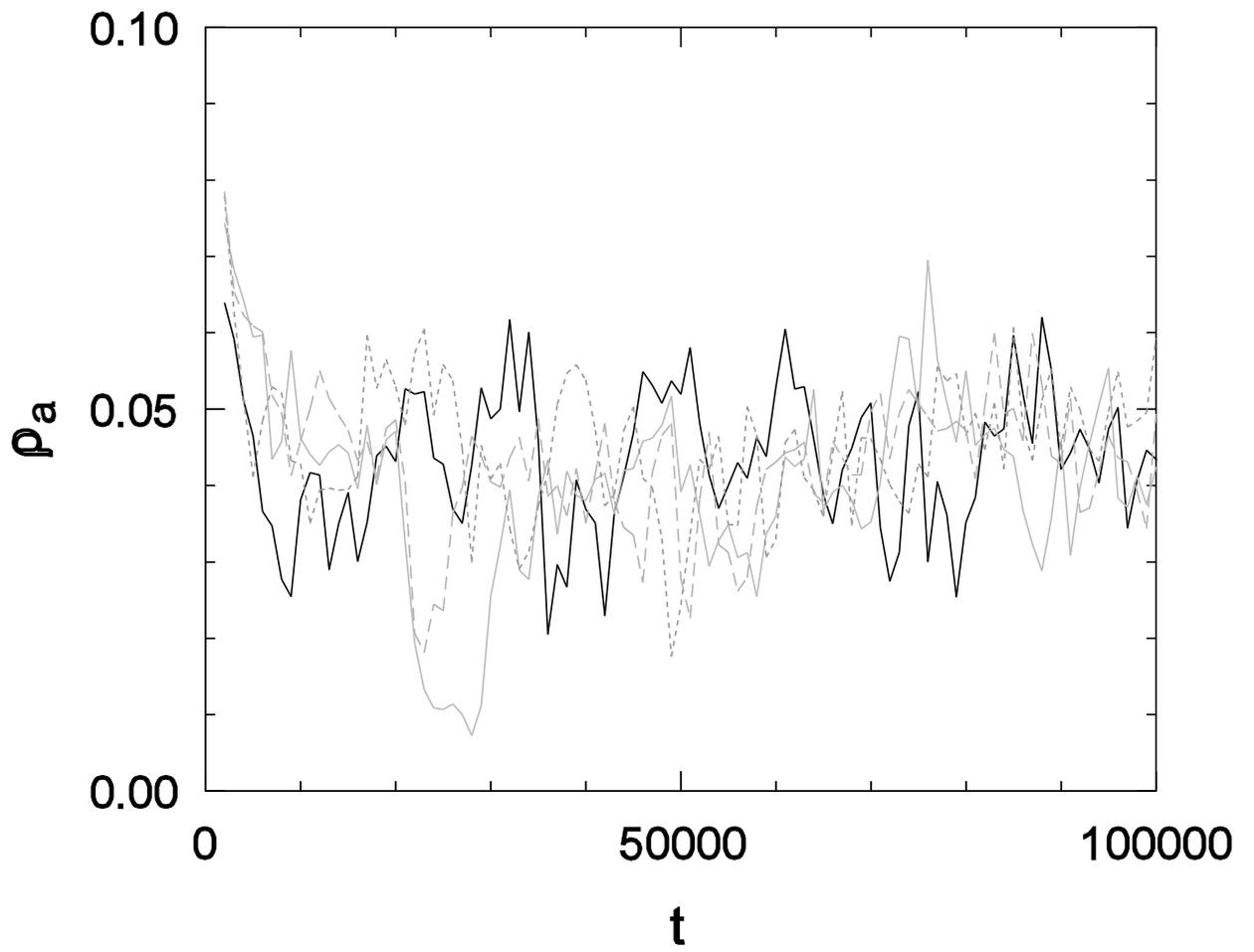
\end{document}